\documentclass{article}

\usepackage{arxiv}

\usepackage{lipsum}		

\usepackage[nolist]{acronym}
\usepackage{subcaption}
\usepackage[ruled]{algorithm2e}
\usepackage{bm} 
\usepackage{natbib}
\usepackage{array, makecell} 
\usepackage{tabularx} 
\usepackage{booktabs}

\usepackage{rotating} 
\usepackage{colortbl} 
\usepackage[dvipsnames, table]{xcolor} 

\usepackage{amsmath}
\usepackage{amssymb}
\usepackage{mathtools}
\usepackage{mathrsfs}

\usepackage{rotating}

\usepackage{geometry}
\usepackage{ragged2e}

\usepackage{setspace}
\usepackage{multirow}
\usepackage{hyperref}
\hypersetup{
    colorlinks,
    citecolor=red,
    filecolor=blue,
    linkcolor=blue,
    urlcolor=blue
}

\usepackage{upgreek}

\newcommand{\0}{\bm{0}}
\newcommand{\1}{\bm{1}}

\newcommand{\bff}{\mathbf{f}}

\newcommand{\bfp}{\mathbf{p}}

\newcommand{\bfA}{\mathbf{A}}

\newcommand{\bftau}{\bm{\uptau}}

\newcommand{\bfomega}{\bm{\upomega}}

\newcommand{\R}{\mathbb{R}}

\DeclarePairedDelimiterX{\Set}[1]{\{}{\}}{ #1}
\DeclarePairedDelimiter{\abs}{\lvert}{\rvert}

\DeclarePairedDelimiterX{\innerp}[2]{\langle}{\rangle}{#1, #2}
\DeclarePairedDelimiterX{\norm}[1]{\lVert}{\rVert}{\ifblank{#1}{\:\cdot\:}{#1}
}

\DeclarePairedDelimiterXPP{\Tr}[1]{\textrm{Tr}}{(}{)}{}{#1}
\DeclarePairedDelimiterXPP{\rank}[1]{\textrm{rank}}{(}{)}{}{#1}
\DeclarePairedDelimiterXPP{\diag}[1]{\textrm{diag}}{(}{)}{}{#1}
\DeclarePairedDelimiterXPP{\vvec}[1]{\textrm{vec}}{(}{)}{}{#1}
\DeclarePairedDelimiterXPP{\mmat}[1]{\textrm{mat}}{(}{)}{}{#1}

\DeclarePairedDelimiterXPP{\Prob}[1]{\mathbb{P}}(){}{  #1}
\DeclarePairedDelimiterXPP{\Expect}[1]{\mathbb{E}}(){}{  #1}
\DeclarePairedDelimiterXPP{\Var}[1]{\text{Var}}(){}{  #1}
\DeclarePairedDelimiterXPP{\Cov}[1]{\text{Cov}}(){}{  #1}

\DeclareMathOperator*{\argmin}{arg\!\,min}

\newcommand{\bffmin}{\bff_{\rm min}}
\newcommand{\bffmax}{\bff_{\rm max}}
\newcommand{\fmin}{f_{\rm min}}
\newcommand{\fmax}{f_{\rm max}}

\newcommand{\bffemg}{\bff_{\rm EMG}}

\newcommand{\bffoc}{\bff_{\rm OC}}

\newcommand{\bfomegaioc}{\bfomega_{\rm IOC}}

\usepackage[normalem]{ulem}

\title{Inverse Optimal Control of Muscle Force Sharing During Pathological Gait}

\author{ \href{https://orcid.org/ 0000-0003-2952-592X}{\includegraphics[scale=0.06]{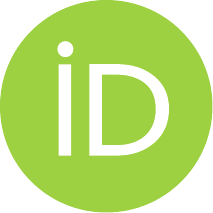}\hspace{1mm}Filip~Be\v{c}anovi\'{c}}\thanks{Corresponding author.} \\
	University of Belgrade \\
	\texttt{filip.becanovic@etf.rs} \\
	\And
	Vincent~Bonnet \\
	LAAS-CNRS \\
	\texttt{vincent.bonnet@laas.fr} \\
	\And
	\href{https://orcid.org/0000-0002-9029-4465}{\includegraphics[scale=0.06]{orcid.pdf}\hspace{1mm}Kosta~Jovanovi\'{c}} \\
	University of Belgrade \\
	\texttt{kostaj@etf.rs} \\
	\And
	\href{https://orcid.org/0000-0001-6738-4529}{\includegraphics[scale=0.06]{orcid.pdf}\hspace{1mm}Samer~Mohamed} \\
	University Paris-Est Creteil \\
	\texttt{samer.mohammed@u-pec.fr} \\
    \And
	\href{https://orcid.org/0000-0002-0368-8248}{\includegraphics[scale=0.06]{orcid.pdf}\hspace{1mm}Rapha\"{e}l~Dumas} \\
	University Gustave Eiffel, Campus Lyon \\
	\texttt{raphael.dumas@eiffel.fr} \\
}

\date{}

\renewcommand{\headeright}{}
\renewcommand{\undertitle}{}
\renewcommand{\shorttitle}{IOC for pathological gait}

\hypersetup{
pdftitle={Inverse Optimal Control of Muscle Force Sharing During Pathological Gait},
pdfsubject={cs.RO, eess.SY, math.OC},
pdfauthor={Filip B.~Be\v{c}anovi\'{c}, Vincent ~Bonnet},
pdfkeywords={Bilevel Optimization and Optimal Control, EMG-driven Muscle Force, Static Optimization, Objective Function, Hemiparesis},
}

\begin{document}
\maketitle

\vspace{-40pt}
\begin{abstract}
    Muscle force sharing is typically resolved by minimizing a specific objective function to approximate neural control strategies. An inverse optimal control approach was applied to identify the "best" objective function, among a positive linear combination of basis objective functions, associated with the gait of two post-stroke males, one high-functioning (subject S1) and one low-functioning (subject S2). It was found that the "best" objective function is subject- and leg-specific. No single function works universally well, yet the best options are usually differently weighted combinations of muscle activation- and power-minimization. Subject-specific inverse optimal control models performed best on their respective limbs (\textbf{RMSE 178/213 N, CC 0.71/0.61} for non-paretic and paretic legs of S1; \textbf{RMSE 205/165 N, CC 0.88/0.85} for respective legs of S2), but cross-subject generalization was poor, particularly for paretic legs. Moreover, minimizing the root mean square of muscle power emerged as important for paretic limbs, while minimizing activation-based functions dominated for non-paretic limbs. This may suggest different neural control strategies between affected and unaffected sides, possibly altered by the presence of spasticity.
Among the 15 considered objective functions commonly used in inverse dynamics-based computations, the root mean square of muscle power was the only one explicitly incorporating muscle velocity, leading to a possible model for spasticity in the paretic limbs. Although this objective function has been rarely used, it may be relevant for modeling pathological gait, such as post-stroke gait.
\end{abstract}

\keywords{Bilevel Optimization and Optimal Control \and EMG-driven Muscle Force \and Static Optimization \and Objective Function \and Hemiparesis}

\section{Introduction} \label{sec:introduction}

In biomechanics, for decades, an optimal solution to muscle force sharing has been obtained by minimizing (or maximizing) a given objective function which is supposed to approximate neural control strategies \citep{bersani2023modeling, prilutsky2002optimization, tsirakos1997inverse}.
The underlying assumption is that the central nervous system efficiently distributes muscle forces to achieve optimal performance. 
A wide range of objective functions have been explored in the literature. While many of these functions have been evaluated \citep{erdemir2007model}, a subset has demonstrated reliability in predicting muscle forces during gait, particularly when validated against electromyography and knee and hip contact force data \citep{modenese2011open, zargham2019inverse}. Based on these evaluations, minimizing a weighted sum of activation or force at a power greater than 2 can be considered a suitable objective function. However, there is no universal agreement on the most effective objective function for accurately predicting muscle forces, especially in the case of pathological gait \citep{falisse2020physics, jansen2014altering, moissenet2013new, serrancoli2014weighted, veerkamp2023predicting, johnson2022patterns, barati2023predictive}.

To estimate muscle force during pathological gait,  EMG-driven  methods (\textit{e.g.}, \citep{li2021well, veerkamp2019effects}), may be currently considered the most efficient way. EMG-driven  methods do not require to define an objective function but require additional instrumentation and complex calibration process. For all the other methods used to predict muscle forces during pathological gait, based on inverse-dynamics or forward dynamics computation \citep{denayer2025prisma, bersani2023modeling}, the definition of the objective function remains an open question.

Within this context, this paper presents an original, data-based, approach to identify the "best" objective function to use for muscle-force sharing during inverse-dynamics computations, particularly one associated with post-stroke gait, precluding the need for EMG-driven estimation. It presents an analysis of many objective functions from the literature, and in particular, presents cross-subject validation of the "best" objective functions identified via inverse optimal control. 
This study relies on EMG-driven muscle estimations \citep{meyer2017lower, li2021well} as ground truth values of muscle forces for learning the "best" objective function weights.
\section{Materials and Methods} \label{sec:methods}

\subsection{Data From Paretic Participants} \label{sec:methods-data}

Published data collected from two poststroke males with right-sided hemiparesis were used for this study. The two participants (S1, height 1.70 m, mass 80.5 kg, age 79 years, and S2, height 1.83 m, mass 88.5 kg, age 62 years,) were described as high functioning (S1, Fugl-Meyer Motor Assessment score of 32 points) and low functioning (S2, Fugl-Meyer Motor Assessment score of 25 points). Quantitative spasticity levels of the subjects were not provided, only S2 was qualitatively assessed as highly-spastic. They walked on a split-belt treadmill at a self-selected speed of 0.5 m/s and 0.45 m/s, respectively. 10 gait cycles were analyzed with a calibrated EMG-driven musculoskeletal model to compute the muscle forces of 35 (for S1) and 34 (for S2) lower limb muscles.
For more details about the data collection, the calibrated musculoskeletal parameters (such as optimal muscle fiber lengths, pennation angles, maximal isometric forces), and the modeling process (including the computation at each sampled instant of time of the joint torques, lever arms, minimal and maximal muscle forces, muscle velocities…), see \citep{li2021well, meyer2017lower}.
All these calibrated musculoskeletal model parameters and variables have been used to formulate the optimal control problem (\textit{i.e.}, definition of objective function, constraints and bounds) predicting the muscle forces during pathological gait.
The state of the musculoskeletal model, through its variables, is used at each sample of time to formulate the inverse dynamics optimization process that models muscle-force sharing.
For example, $\bffmin$ and $\bffmax$ were previously obtained at each sampled instant of time using the EMG-driven musculoskeletal model. 
Also, note that the calibrated musculoskeletal model assumed a rigid tendon, therefore the muscle velocity refers to fiber velocity.
To estimate additional musculoskeletal parameters, such as muscle physiological cross-sectional areas and muscle masses, a specific tension of 61 N/cm² and a density of 1059.7 kg/m³ were assumed.
\subsection{Optimal Control} \label{sec:static-optimization}

When muscle force sharing is not solved by a calibrated EMG-driven musculoskeletal model as in \citep{meyer2017lower}, the muscle forces are classically computed, at each sampled instant of time by an inverse-dynamics based computation:
\begin{subequations} \label{eq:force-sharing-optimization}
\begin{align}
    \bffoc = \argmin_{\bff \in \R^{n_m}} \quad & \Phi(\bff; \bfp) \label{eq:force-sharing-optimization-objective} \\
    \text{subject to} \quad & \bftau = 
    \underbrace{ \bfA 
    \begin{bmatrix}
        \ddots &  & \\
        & \cos(\alpha^j) & \\
        & & \ddots
    \end{bmatrix}
    }_{\textstyle \tilde \bfA} 
    \bff \label{eq:force-sharing-optimization-sharing-equation} \\
    & \bffmin \preceq \bff \preceq \bffmax \label{eq:force-sharing-optimization-limits}
\end{align}
\end{subequations}

with $\bff$ the vector of muscle forces, $\Phi$ a chosen objective function, $\bfp$ the model parameters and variables obtained in the calibration of the EMG-driven musculoskeletal model including $\bftau$ the vector of joint torques, $\bfA$ the matrix of lever arms, $\alpha^j$ the pennation angle of muscle $j$ (introduced in a diagonal matrix), $\bffmin$ and $\bffmax$ the vectors of minimal and maximal muscle forces.

The vector $\bff$ indeed refers to the muscle-fiber force as opposed to the term “musculo-tendon force”, classically used when no hill type model is introduced. It is important to note that many of these variables (\textit{e.g.}, joint torques $\bftau$, muscle moment arms $\bfA$, pennation angles $\alpha^j$) were obtained at each sampled instant of time from the state of the EMG-driven musculoskeletal model \citep{li2021well, meyer2017lower}.

This optimization corresponds to a direct optimal control approach, where the optimal muscle forces $\bffoc$ minimize an objective function $\Phi$ that is postulated to model the central nervous system's strategy for muscle force sharing. This direct optimal control approach applies to an inverse dynamics-based computation as the joint torques are derived from motion capture data. The term ‘direct’ implies muscle forces are computed given an arbitrary objective function. 15 objective functions commonly used in biomechanics for muscle force sharing in the context of inverse-dynamics computations have been identified in the literature \citep{praagman2006relationship, rasmussen2001muscle, tsirakos1997inverse} and listed in Table \ref{tab:basis-objectives}. Power objective functions were transformed into norms, using square or cubic roots, preserving convexity and the optimal solution, while homogenizing the sensitivity of the objective functions $\phi_i$ to the muscle forces $\bff$ \citep{boyd2004convex}. 
Objective functions $\phi_4$ -- $\phi_8$ contain terms depending on muscle activations, which are expressed as normalized forces: $a^{j} = \frac{f^j - \fmin^j}{\fmax^j - \fmin^j}$, where $a^{j}$ denotes the activation of muscle $j$, $\fmin^j$ represents the passive muscle force (corresponding to zero activation), and $f^j - \fmin^j$ represents the active muscle force.

To be able to compare the results of this optimization with the muscle forces previously obtained using the EMG-driven musculoskeletal model (\S\ref{sec:methods-data}), the exact same model parameters and variables at each sampled instant of time are used as previously proposed by  \citep{de2016evaluation} to compare direct collocation and static optimization.
Specifically, parameters and variables of the EMG-driven musculoskeletal model \citep{li2021well, meyer2017lower} have been used for optimal control \eqref{eq:force-sharing-optimization} in the following ways:
\begin{enumerate}
    \item torques, muscle moment arms, and pennation angles are used explicitly in \eqref{eq:force-sharing-optimization-sharing-equation},
    \item minimal (\textit{i.e.,} passive) muscle force and maximal muscle forces are used explicitly in \eqref{eq:force-sharing-optimization-limits},
    \item torques, minimal muscle forces, maximal muscle forces, physiological cross-sectional areas, muscle fiber velocities, maximal joint torques generated by single muscles, maximal isometric forces, muscle masses are used implicitly in the objective function \eqref{eq:force-sharing-optimization-objective} and their use is defined in Table \ref{tab:basis-objectives}.
\end{enumerate} 
By re-using the state of the musculoskeletal model at each time-instant, contained in the value of these parameters and variables, we compute the forces $\bffoc$ in an inverse-dynamics computation, and use inverse optimal control (\S\ref{sec:methods-inverse}) to identify the most accurate weights (\textit{i.e.,} "best") one should use when inferring muscle forces through so called static optimization.

\subsection{Inverse Optimal Control} \label{sec:methods-inverse}

If the muscle force sharing is solved by a calibrated EMG-driven musculoskeletal model \citep{meyer2017lower}, the muscle forces $\bffemg$ are computed without choosing any objective function. In this case, the Inverse Optimal Control (IOC) approach involves identifying the objective function that generates the closest muscle forces when minimized using inverse-dynamics based computation  \citep{bevcanovic2023force}. The identification is formulated as the minimization of the mean-squared error between the EMG-driven (\textit{i.e.}, observed) muscle forces $\bffemg$ and the muscle forces obtained by inverse-dynamics computation (i.e., optimal) $\bffoc$ obtained with a weighted objective function: 
\begin{subequations} \label{eq:ioc}
\begin{align}
    \bfomegaioc = \argmin_{\bfomega \in \R^m} \quad & \sum_{d=1}^{\abs{\mathcal{D}}} \norm{\bffemg^{(d)} - \bffoc^{(d)}}_2^2 \label{eq:ioc-objective} \\
    \text{subject to} \quad & \bfomega \succeq \0 \label{eq:ioc-positivity} \\
    & \1^T \bfomega = 1 \label{eq:ioc-regularity} \\
    &
    \begin{aligned}[t]
    \bffoc^{(d)} = \argmin_{\bff \in \R^{n_m}} \quad & \sum_{i=1}^m \omega_i \phi_i(\bff; \bfp^{(d)}) \\
    \text{subject to} \quad & \bftau^{(d)} =
    \tilde\bfA^{(d)} 
    \bff \\
    & \bffmin^{(d)} \preceq \bff \preceq \bffmax^{(d)}
    \end{aligned}
    \label{eq:ioc-innerloop}
\end{align}
\end{subequations}
with $\bfomega$ a vector of weights and $\norm{}_2^2$ the mean-squared error between the observed and the optimal muscle forces at each point $d$ of a data set $\mathcal{D}$ (\textit{e.g.}, at each sampled instant of time of the gait cycle).

\begin{figure}[!ht]
    \caption{}
    \label{fig:direct-inverse}
    \centering
    \includegraphics[width=\linewidth]{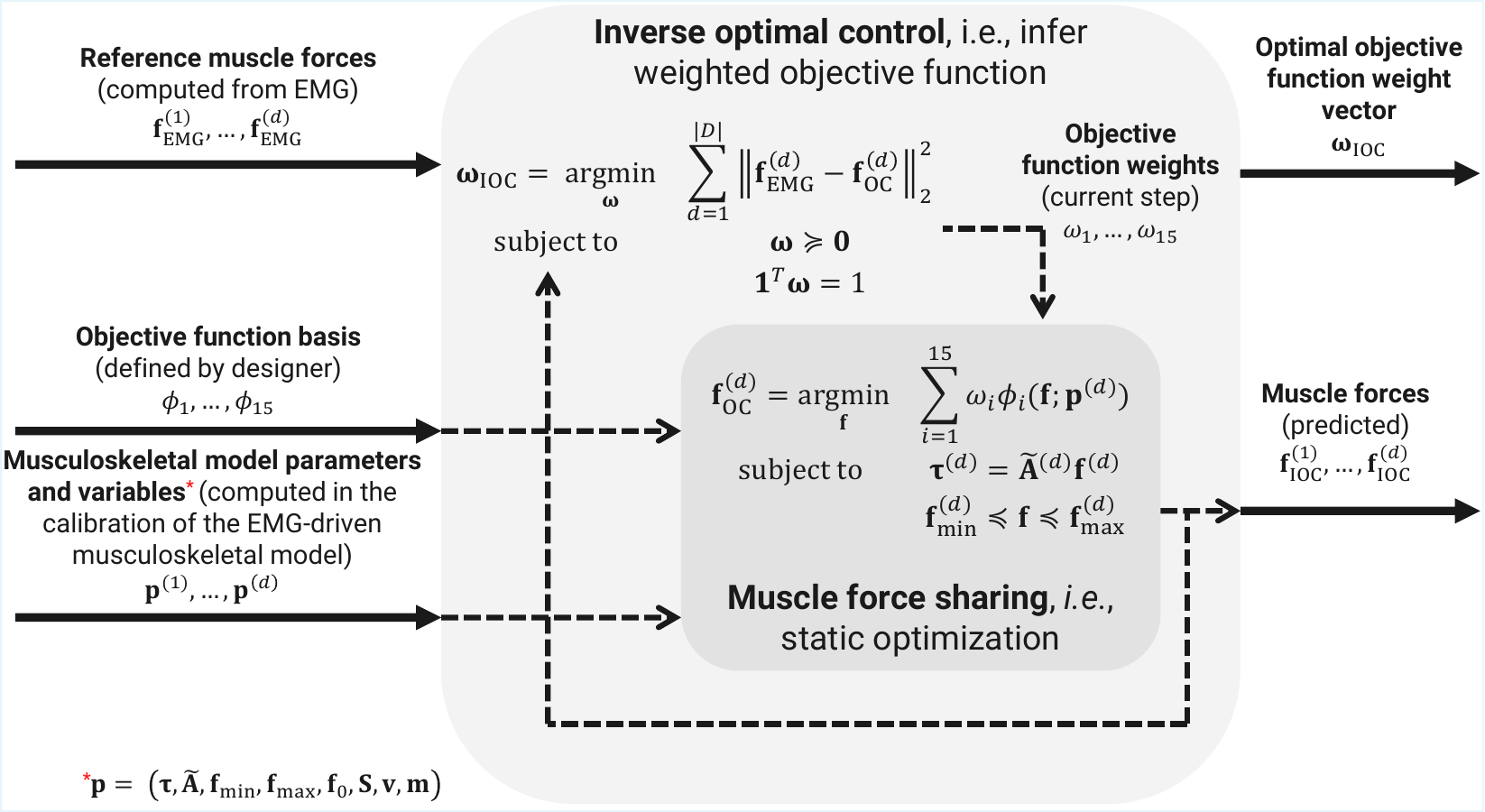}
\end{figure}

Four different bilevel optimizations of the weight vector $\bfomega$ (Figure \ref{fig:direct-inverse}) were performed on each participant's 10 available gait cycles of both paretic and non-paretic legs. The corresponding four identified objective functions, \textit{i.e.,} $\Phi^{\rm S1, LEG1}_{\rm IOC}$, $\Phi^{\rm S1, LEG2}_{\rm IOC}$, $\Phi^{\rm S2, LEG1}_{\rm IOC}$, and $\Phi^{\rm S2, LEG2}_{\rm IOC}$, are reported through their corresponding weights, \textit{i.e.,} $\bfomega^{\rm S1, LEG1}_{\rm IOC}$, $\bfomega^{\rm S1, LEG2}_{\rm IOC}$, $\bfomega^{\rm S2, LEG1}_{\rm IOC}$, and $\bfomega^{\rm S2, LEG2}_{\rm IOC}$, within Table \ref{tab:ioc-weights}. The root-mean-squared error (RMSE) and correlation coefficient (CC) between the observed ($\bffemg$) and the optimal muscle forces ($\bffoc$) was calculated for each of the 15 objective functions from Table \ref{tab:basis-objectives} (with $\omega_i = 1$ in \eqref{eq:force-sharing-optimization-objective}) and with the results of inverse optimal control (with $\bfomega = \bfomega_{\rm IOC}$ in \eqref{eq:force-sharing-optimization-objective}) and will be reported in Table \ref{tab:cross-validation}.

Since, in non-convex bilevel problems, the initial point can influence the final solution, we ensure the solution of \eqref{eq:ioc} is not a local minimum by performing a systematic sweep through the space $\R^m$ of $\bfomega$ (grid-points) and initializing a local-search from there. This accounts for a more robust convergence process.
Moreover, to ensure that the outer problem is well-conditioned—\textit{i.e.}, that the sensitivity of the inner-loop solution with respect to each objective function weight is comparable—we normalized each objective function by its maximum value over the dataset: $\phi_j := \phi_j / {\phi_j}_{\rm max}$.
To solve the inner-loop \eqref{eq:ioc-innerloop}, we used CasADi \citep{Andersson2018} coupled with IPOPT, while we used MATLAB's \texttt{fmincon} to perform the local-search for $\bfomega$ in \eqref{eq:ioc}, with all runs stopping at an actual local minimum (with default tolerances). 
For each of the four runs, the data consisted of 10 recorded gait cycles with 101 samples per cycle, yielding a data set $\mathcal{D}$ size of $|\mathcal{D}| = 1010$ in \eqref{eq:ioc-objective}.
The computations related to each run lasted between 5 and 10 hours.

Cross-validation between subjects is performed by identifying an objective function, via inverse optimal control, on the force data of a given leg of a given subject and then testing it against the observed force data of the corresponding leg of the other subject.
We assumed that muscle distribution processes within the non-paretic leg and the paretic one are driven by different objective functions.
For the subject--leg--specific weighted  objective fucntion $\Phi_{\rm IOC}$ (bilevel-identified weights), we report RMSE and CC on held-out data.

\section{Results} \label{sec:results}

Table \ref{tab:ioc-weights} presents the optimal weight vectors $\bfomegaioc$ obtained from the four inverse optimal control runs, one for each subject-leg pair. 
Seven individual objective functions were associated with null weight in all cases: $\phi_1$ (average muscle force), $\phi_2$ (root mean square of muscle force), $\phi_3$ (cubic average of muscle force), $\phi_7$ (cubic average of muscle activation), $\phi_9$ (average muscle stress), $\phi_{10}$ (root mean square of muscle stress), and $\phi_{12}$ (maximum muscle stress), and their columns are greyed out in the table. Five objective functions were associated with low contributions ($\omega \leq 0.1$) to overall weighting in all cases: $\phi_4$ (maximum muscle force), $\phi_5$ (average muscle activation), $\phi_{11}$ (cubic average of muscle stress), $\phi_{14}$ (root-mean square of torque-normalized muscle force), and $\phi_{15}$ (metabolic energy), and their columns are highlighted in yellow in the table. Meanwhile, three objective functions stand out with their strong contribution ($\omega \geq 0.1$) to overall weighting in some of the four cases: $\phi_6$ (root mean square of muscle activation), $\phi_8$ (maximum muscle activation), and $\phi_{13}$ (root mean square of muscle power), and their columns are highlighted in green in the table.

Objective functions with high values for the weight can be considered representative of the neural control strategies. However, they are not consistent between the legs and the participants. Yet, high weight values for activation-based objective functions ($\phi_6$ - root mean square of muscle activation, and $\phi_8$ - maximum muscle activation) emerge for the muscle force sharing of the non-paretic leg ($\rm LEG1$) with $\omega_8 = 0.94$ for S1 and $\omega_6 = 0.47$ for S2. On the other hand, high weight values for the power-based objective function ($\phi_{13}$ - root mean square of muscle power) emerge for the muscle force-sharing of the paretic leg ($\rm LEG2$) with $\omega_{13} = 0.56$ for S1 and $\omega_{13} = 0.9$ for S2. 

Table \ref{tab:cross-validation} shows cross-validation of subject--leg--specific weighted objective function across two participants, reported separately for the non-paretic (green) and paretic (yellow) legs.
Rows $\Phi^{\rm S1, LEG1}_{\rm IOC}$, $\Phi^{\rm S1, LEG2}_{\rm IOC}$, $\Phi^{\rm S2, LEG1}_{\rm IOC}$, and $\Phi^{\rm S2, LEG2}_{\rm IOC}$ are subject--leg--specific weighted objective function (IOC-identified weights), evaluated where meaningful (cells that are empty are not applicable). The \textit{Rank (left)} columns give the within-block rank of the adjacent metric (1 = best; lower RMSE / higher CC is better). In all four subject--leg blocks, the weighted objective function learned for that leg achieve rank 1 (\textit{e.g.,} S1 non-paretic: RMSE = 178 N, CC = 0.71; S2 non-paretic: 205 N, 0.88; S1 paretic: RMSE = 213 N, CC = 0.61; S2 paretic: RMSE = 165 N, CC = 0.85) and transfer reasonably across subjects (ranks 2--4). Among single-term candidates, $\phi_6$--$\phi_7$ are strongest on non-paretic legs (S1: RMSE = 187 N, CC = 0.66; S2: RMSE = 208--214 N, CC = 0.87), while $\phi_4$ (S1 paretic: RMSE = 236 N, CC = 0.44) and $\phi_{13}$ (S2 paretic: RMSE = 181 N, CC = 0.83) are the best single terms on the paretic side. Overall, subject--leg--specific weighted objective functions consistently outperform any single-term objective, with clearer gains on the paretic leg.

The fact that an objective function gives good predictions by itself (\textit{e.g.} $\phi_4$ for S1 paretic -- RMSE $236$ N (Rank 2), CC $0.44$ (Rank 6) -- in Table \ref{tab:cross-validation}) does not necessarily mean it will have a strong presence in the identified weight vector ($\bfomega^{\rm S1, LEG2}_{\rm IOC}$ -- $\omega_4 = 0.01$ -- in Table \ref{tab:ioc-weights}). Nevertheless, it is often the case that when an objective function performs well by itself (\textit{e.g.,} $\phi_6$--$\phi_7$ on S1 non-paretic -- RMSE $187$ N (Rank 3), CC $0.66$ (Rank 3) -- and S2 non-paretic -- RMSE $208$ N (Rank 2), CC $0.87$ (Rank 2) -- in Table \ref{tab:cross-validation}), it, or an objective function similar to it (\textit{i.e.,} $\phi_8$), will have a strong presence in the identified weight vector ($\bfomega^{\rm S1, LEG1}_{\rm IOC}$ -- $\omega_8 = 0.94$ -- and $\bfomega^{\rm S2, LEG1}_{\rm IOC}$ -- $\omega_6 = 0.47$, $\omega_8 = 0.17$ -- in Table \ref{tab:ioc-weights}).

A modest generalization capacity is observed in the non-paretic leg across subjects. The subject-specific IOC solutions demonstrate this limited transferability: $\Phi^{\rm S1, LEG1}_{\rm IOC}$ achieves reasonable performance when applied to S2's non-paretic leg (RMSE = 216 N, CC = 0.86, rank 4), while $\Phi^{\rm S2, LEG1}_{\rm IOC}$ maintains competitive performance on S1's non-paretic leg (RMSE = 184 N, CC = 0.68, rank 2). However, neither cross-subject application achieves the optimal performance of the subject-specific solution, indicating that even non-paretic limbs retain individual motor control characteristics that limit complete generalization.

The paretic legs demonstrate a complete lack of generalization capacity, particularly when transferring models from the less functional to the more functional subject. $\Phi^{\rm S2, LEG2}_{\rm IOC}$, optimized for S2's paretic leg, shows dramatically poor performance when applied to S1's paretic leg (RMSE = 283 N, CC = 0.31, rank 15), representing one of the worst-performing models for this condition. Conversely, $\Phi^{\rm S1, LEG2}_{\rm IOC}$ performs moderately when applied to S2's paretic leg (RMSE = 257 N, CC = 0.57, rank 3), but still substantially underperforms the subject-specific solution ($\Phi^{\rm S2, LEG2}_{\rm IOC}$: RMSE = 165 N, CC = 0.85, rank 1).

The muscle power-based objective function ($\phi_{13}$) exhibits distinctly different behavior patterns between subjects and legs. For S2 (less functional), $\phi_{13}$ demonstrates exceptional performance on the paretic leg (RMSE = 181 N, CC = 0.83, rank 2), nearly matching the subject-specific IOC solution. However, this same function performs poorly across all other conditions: S1's non-paretic leg (RMSE = 276 N, CC = 0.29, rank 17), S1's paretic leg (RMSE = 303 N, CC = 0.27, rank 16), and S2's non-paretic leg (RMSE = 386 N, CC = 0.48, rank 17).

This selective high performance specifically for the less functional subject's pathological leg may suggest that the muscle power-based objective function effectively captures the motor control strategy associated with post-stroke spasticity.
\section{Discussion} \label{sec:conclusion}
The objective of this study was to identify the ”best” objective function to use for muscle-force sharing in post-stroke gait using inverse optimal control.
Inverse optimal control is generally used to infer optimality principles from observed trajectories. It can be applied to the muscle forces estimated with a calibrated EMG-driven musculoskeletal model to identify the objective function generating the closest muscle forces \citep{bevcanovic2023force}.
Representing a human subject's objective function as a linear combination of basis objective functions, as in Equation \eqref{eq:ioc-innerloop}, is a standard approach in inverse optimal control literature \citep{chan2025inverse}. While non-linear combinations represent a potentially valuable avenue for investigation, the limited prior research on such formulations motivated our choice of the linear framework. Furthermore, the selection of mean-squared error minimization for identifying cost function weights, rather than alternative loss functions such as the $L_1$ or $L_\infty$ norms, is consistent with common practice in the field \citep{chan2025inverse}. This choice inherently biases the inverse optimal control weight identification toward fitting larger muscles that produce greater forces, whereas alternative loss functions would yield different prioritizations.

It is important to note that a calibrated EMG-driven musculoskeletal model does not involve any optimization other than calibration of the model parameters (such as optimal muscle fiber lengths, pennation angles, and maximal isometric forces) to match the joint torques derived from motion capture data \citep{meyer2017lower, hammond2025neuromusculoskeletal}. These muscle forces derived from EMG were considered the observed ones. It was thus decided to formulate the optimization problem with muscle forces, not musculo-tendon forces. This way, calculating the activations and the active muscle forces involved in the 15 objective functions tested was straightforward.

The 15 objective functions tested are classically used for inverse-dynamics-based computations, with some of them being attributed physical meaning: like $\phi_7$, which has been linked to fatigue, or $\phi_{15}$, which has been linked to metabolic energy. They only involve muscle forces or activations but simultaneous minimization of ligament and joint contact forces was also proposed \citep{hu2013influence, moissenet20143d}. Other objective functions, such as the rate of metabolic energy, are additionally used when forward-dynamics-based computations are used \citep{febrer2023predictive}. Previous evaluation or comparison of objective functions were more generally performed in this forward dynamics simulation scheme \citep{ackermann2010optimality, nguyen2019bilevel, tomasi2023identification, zargham2019inverse}. 

The results of inverse optimal control confirm that the identified objective functions for each individual achieve the best performance for their specific subject and leg, validating a subject-leg-specific approach for muscle force prediction during inverse dynamics computations.
In the present study, the objective functions involving muscle activation and muscle power were identified as optimal while the objective functions involving muscle force or muscle stress were not. Minimizing the cubic average of muscle activation has been previously reported to provide reliable results \citep{ackermann2010optimality, zargham2019inverse}. Minimizing the root mean square of muscle power has not been widely used or evaluated in the past but seems to be relevant for post-stroke gait. This objective function performed specifically for the less functional subject's pathological leg. This suggests that muscle power may represent the motor control strategy associated with post-stroke spasticity. This is consistent with the modeling of spasticity previously introduced in different musculoskeletal models for simulating post-stroke gait \citep{jansen2014altering, moissenet2013new, veerkamp2023predicting}.
However, it is important to note that the weighed objective function identified by inverse optimal control may not only reflect motor control but also biomechanical affordances (i.e., musculoskeletal structure) as well as biomechanical constraints (i.e., task requirements). The root mean square of muscle power is the only objective function that explicitly involves muscle velocity, which is necessarily minimized in spastic gait.

The results of the inverse optimal control presented limited cross-subject generalization, even though some degree of it was observed for the non-paretic leg.  Transferring the identified objective function from the less functional (qualitatively assessed as highly-spastic) to the more functional subject revealed the less efficient. Objective functions like $\phi_{13}$ (square root of muscle power) outperformed generalized IOC solutions, suggesting that certain universal pathological motor patterns may be more effectively captured by appropriately selected objective functions than by cross-subject transfer of identified ones. Current findings indicate that it is not probable to identify, by combining the 15 selected ones, a general objective function that will perform close-to-optimally for many subjects at once.

This study presents several limitations. First, the sample size ($n=2$) is extremely limited, in part because of the difficulty in acquiring data of this quality, restricting the generalization of the findings. The two participants were high- and low-functioning, and different objective functions were identified for the four legs studied; however, quantitative levels of muscle spasticity or contracture were not available. High-quality data from a healthy control group would have been useful to draw a clearer line between the non-paretic and paretic leg control strategies. Second, the observed muscle forces are derived from the calibrated EMG-driven musculoskeletal instead of directly measured, which is impossible in vivo. Third, the optimal objective function is selected from a limited non-independent set and considers an entire movement phase. These are acknowledged issues for inverse optimal control. Moreover, it can be hypothesized that this strategy is not optimal but suboptimal in pathological gait \citep{bersani2023modeling}. Actually, the distances between the observed and the optimal muscle forces were generally higher for the paretic legs of the participants.

\section{Conclusion}

In conclusion, the inverse optimal control approach was effectively applied to muscle forces within an inverse dynamics-based computation. The muscle forces computed by optimizing the IOC-identified objective functions were compared to observed muscle forces  obtained with a calibrated EMG-driven musculoskeletal model. Minimizing the root mean square- and maximum-activation, and the root mean square of muscle power were the three main objective functions identified  by the inverse optimal control procedure. The paretic leg of the less functional subject demonstrated a specific optimal (or more probably non-optimal) control and cross-subject generalization was limited. Although the cross-subject generalization of inverse optimal control may be not achieved, the study has identified common patterns for the "best" objective function across subjects and provides recommendations on choosing the objective function for pathological gait in inverse-dynamics based computation. Typically, objective functions involving muscle force or muscle stress were not identified as optimal. The proposed approach could be useful to further investigate the central nervous system's strategy for muscle force sharing in pathological subjects.

Future work should extend this approach to larger cohorts to identify generalizable trends in optimal objective function selection, potentially stratified by pathology severity or spasticity level. Such investigations would necessitate the acquisition of extensive neuromechanical datasets from both research and clinical settings, as well as their dissemination to the broader scientific community.
\clearpage
\section*{Figure Captions}

\textbf{Table \ref{tab:basis-objectives}}: Basis objective functions commonly used in biomechanics for inverse dynamics-based muscle force sharing.

\vspace{12pt} \noindent
\textbf{Table \ref{tab:ioc-weights}}:
Recovered IOC weight vectors ($\omega_1$ - $\omega_{15}$) for two subjects (S1, S2) and both legs (LEG1, LEG2); rows are approximately normalized, solutions are sparse, and column shading groups feature families for visual reference.

\vspace{12pt} \noindent
\textbf{Table \ref{tab:cross-validation}}: 
Cross-validation table of inverse-optimal muscle force sharing predictions across two subjects (S1 and S2), reporting Root Mean Squared Error (RMSE), Pearson's Correlation Coefficient (CC), and ranking for both functional and paretic legs.

\vspace{12pt} \noindent
\textbf{Figure 1}: 
Principle of inverse optimal control applied to muscle force sharing.

\vspace{12pt} \noindent
\textbf{Supplementary Figure A}: Muscle forces from the data set and those obtained from inverse optimal control for a selected gait cycle (\#10 out of 10).
The figure depicts the data set medians (blue lines; $f_{50\%}$), the 25\textsuperscript{th}-to-75\textsuperscript{th} percentile regions (blue-shaded areas; $[f_{25\%}, f_{75\%}]$), and total range regions (gray-shaded areas; $[f_{0\%}, f_{100\%}]$) of the muscle forces $f_{\rm EMG}$ across the 10 gait cycles for each of the 35 investigated muscles \citep{meyer2017lower}. Also depicted are the muscle forces (dashed green lines; $f_{\rm EMG}$) of one particular gait cycle, the 10\textsuperscript{th} one out of 10 in total \citep{meyer2017lower}, the proposed prediction using the inversely optimal objective function (dotted pink lines; $f_{\rm IOC}$), and the proposed cross-validation using the inversely optimal objective function identified on the other subject's data (dash-dotted red lines; $f_{\rm CV}$).

The depicted muscle force trajectories $f^j$ correspond to muscles: $j=1$ adductor brevis; $j=2$ adductor longus; $j=3$ adductor magnus distal; $j=4$ adductor magnus ischial; $j=5$ adductor magnus middle; $j=6$ adductor magnus proximal; $j=7$ gluteus maximus superior; $j=8$ gluteus maximus middle; $j=9$ gluteus maximus inferior; $j=10$ gluteus medius anterior; $j=11$ gluteus medius middle; $j=12$ gluteus medius posterior; $j=13$ gluteus minimus anterior; $j=14$ gluteus minimus middle; $j=15$ gluteus minimus posterior; $j=16$ iliacus; $j=17$ psoas; $j=18$ semimembranosus; $j=19$ semitendinosus; $j=20$ biceps femoris long; $j=21$ biceps femoris short; $j=22$ rectus femoris; $j=23$ vastus medialis; $j=24$ vastus intermedius; $j=25$ vastus lateralis; $j=26$ lateral gastronemius; $j=27$ medial gastronemius; $j=28$ tibialis anterior; $j=29$ tibialis posterior; $j=30$ peroneus brevis; $j=31$ peroneus longus; $j=32$ peroneus tertius; $j=33$ soleus; $j=34$ extensor digitorum longus; $j=35$ flexor digitorum longus.

\clearpage

\begin{table}[!ht]
    \caption{Basis objective functions commonly used in biomechanics for inverse dynamics-based muscle force sharing}
    \label{tab:basis-objectives}
    \centering
    \begin{tabularx}{\textwidth}{|c|c|X|}
        \hline
        Number & Objective function & Description \\ \hline \hline
        $\phi_1$ & $\displaystyle \frac{1}{{n_m}} \sum_{j=1}^{n_m} f^j$ & Minimize the average muscle force \\
        $\phi_2$ & $\displaystyle \left( \frac{1}{{n_m}} \sum_{j=1}^{n_m} (f^j)^2 \right)^{\frac{1}{2}}$ & Minimize the root mean square of muscle force \\
        $\phi_3$ & $\displaystyle \left( \frac{1}{{n_m}} \sum_{j=1}^{n_m} (f^j)^3 \right)^\frac{1}{3}$ & Minimize the cubic average of muscle force \\
        $\phi_4$ & $\displaystyle \max_{j = 1, \ldots, {n_m}}{f^j}$ & Minimize the peak muscle force \\
        $\phi_5$ & $\displaystyle \frac{1}{{n_m}} \sum_{j=1}^{n_m} \frac{f^j - \fmin^j}{\fmax^j - \fmin^j}$ & Minimize the average muscle activation \\
        $\phi_6$ & $\displaystyle \left( \frac{1}{{n_m}} \sum_{j=1}^{n_m} \left( \frac{f^j - \fmin^j}{\fmax^j - \fmin^j} \right)^2 \right)^\frac{1}{2}$ & Minimize the root mean square of muscle activation \\
        $\phi_7$ & $\displaystyle \left( \frac{1}{{n_m}} \sum_{j=1}^{n_m} \left( \frac{f^j - \fmin^j}{\fmax^j - \fmin^j} \right)^3 \right)^\frac{1}{3}$ & Minimize the cubic average of muscle activation \\
        $\phi_8$ & $\displaystyle \max_{j = 1, \ldots, {n_m}}{ \frac{f^j - \fmin^j}{\fmax^j - \fmin^j} }$ & Minimize the peak of muscle activation \\
        $\phi_9$ & $\displaystyle \frac{1}{{n_m}} \sum_{j=1}^{n_m} \frac{f^j}{S^j}$ & Minimize the average muscle stress \\
        $\phi_{10}$ & $\displaystyle \left( \frac{1}{{n_m}} \sum_{j=1}^{n_m} \left( \frac{f^j}{S^j} \right)^2 \right)^\frac{1}{2}$ & Minimize the root mean square of muscle stress \\
        $\phi_{11}$ & $\displaystyle \left( \frac{1}{{n_m}} \sum_{j=1}^{n_m} \left( \frac{f^j}{S^j} \right)^3 \right)^\frac{1}{3}$ & Minimize the cubic average of muscle stress \\
        $\phi_{12}$ & $\displaystyle \max_{j = 1, \ldots, {n_m}}{ \frac{f^j}{S^j} }$ & Minimize the peak muscle stress \\
        $\phi_{13}$ & $\displaystyle \left( \frac{1}{{n_m}} \sum_{j=1}^{n_m} \left( f^j v^j \right)^2 \right)^\frac{1}{2}$ & Minimize the root mean square of muscle power \\
        $\phi_{14}$ & $\displaystyle \left( \frac{1}{{n_m}} \sum_{j=1}^{n_m} \left( \frac{f^j}{\norm{\bftau_{\rm max}^j}_1} \right)^2 \right)^\frac{1}{2}$ & Minimize the root mean square of muscle force normalized by joint torque \\
        $\phi_{15}$ & $\displaystyle \Bigg( \frac{1}{{n_m}} \sum_{j=1}^{n_m} \frac{{\rm m}^j}{2} \Bigg( \frac{f^j - \fmin^j}{f_0^j} + \left( \frac{f^j - \fmin^j}{S^j} \right)^2 \Bigg) \Bigg)^\frac{1}{2}$ & Minimize metabolic energy (related to normalized active muscle force and active muscle stress) \\ 
        \hline
    \end{tabularx}
\justifying
\par 
\noindent
where $n_m$ is the number of muscles, $\frac{f^j - \fmin^j}{\fmax^j - \fmin^j} = a^{j}$ the activation of muscle $j$, $S^j$ the physiological cross-sectional area of muscle $j$, $v^j$ the velocity of muscle $j$,  $\bftau_{\rm max}^j$ the maximal joint torque generated by muscle $j$, $m^j$ the mass of muscle $j$, and $f_0^j$ the isometric force of muscle $j$.  \underline{\textbf{Note:}} $\bftau_{\rm max}^j = \tilde\bfA \cdot (0, \ldots, 0, f^j_{\rm max}, 0, \ldots, 0)$ and $ \norm{\bftau_{\rm max}^j}_1 = \sum_{k} \tau_{\rm max, k}^j $.
\end{table}

\begin{table}[!ht]
    \centering
    \caption{}
    \label{tab:ioc-weights}
    \begin{tabular}{|>{\columncolor{gray!10}}l|>{\columncolor{gray!50}}l|>{\columncolor{gray!50}}l|>{\columncolor{gray!50}}l|>{\columncolor{yellow!20}}l|>{\columncolor{yellow!20}}l|>{\columncolor{green!10}}l|>{\columncolor{gray!50}}l|>{\columncolor{green!10}}l|>{\columncolor{gray!50}}l|>{\columncolor{gray!50}}l|>{\columncolor{yellow!20}}l|>{\columncolor{gray!50}}l|>{\columncolor{green!10}}l|>{\columncolor{yellow!20}}l|>{\columncolor{yellow!20}}l|}
    \hline
    \rowcolor{gray!10}
        ~ & $\omega_1$ & $\omega_2$ & $\omega_3$ & $\omega_4$ & $\omega_5$ & $\omega_6$ & $\omega_7$ & $\omega_8$ & $\omega_9$ & $\omega_{10}$ & $\omega_{11}$ & $\omega_{12}$ & $\omega_{13}$ & $\omega_{14}$ & $\omega_{15}$ \\ \hline
        \rule{0pt}{15pt}
        $\bfomega^{\rm S1, LEG1}_{IOC}$ & 0 & 0 & 0 & 0.01 & 0 & 0.01 & 0 & 0.94 & 0 & 0 & 0.01 & 0 & 0.02 & 0.01 & 0 \\ \hline
        \rule{0pt}{15pt}
        $\bfomega^{\rm S2, LEG1}_{IOC}$ & 0 & 0 & 0 & 0 & 0.1 & 0.47 & 0 & 0.17 & 0 & 0 & 0 & 0 & 0.27 & 0 & 0 \\ \hline
        \rule{0pt}{15pt}
        $\bfomega^{\rm S1, LEG2}_{IOC}$ & 0 & 0 & 0 & 0.05 & 0 & 0.34 & 0 & 0 & 0 & 0 & 0 & 0 & 0.56 & 0.05 & 0 \\ \hline
        \rule{0pt}{15pt}
        $\bfomega^{\rm S2, LEG2}_{IOC}$ & 0 & 0 & 0 & 0.01 & 0 & 0 & 0 & 0.05 & 0 & 0 & 0 & 0 & 0.9 & 0 & 0.04 \\ \hline
    \end{tabular}
\end{table}
\begin{sidewaystable}[!ht]
    \centering
    \caption{Cross-validation table}
    \label{tab:cross-validation}
    \begin{tabular}{|>{\columncolor{gray!10}}c|>{\columncolor{green!10}}c|>{\columncolor{green!10}}c|>{\columncolor{green!10}}c|>{\columncolor{green!10}}c|>{\columncolor{yellow!20}}c|>{\columncolor{yellow!20}}c|>{\columncolor{yellow!20}}c|>{\columncolor{yellow!20}}c|>{\columncolor{green!10}}c|>{\columncolor{green!10}}c|>{\columncolor{green!10}}c|>{\columncolor{green!10}}c|>{\columncolor{yellow!20}}c|>{\columncolor{yellow!20}}c|>{\columncolor{yellow!20}}c|>{\columncolor{yellow!20}}c|}
    \hline
        \rowcolor{gray!10}
        ~ & \multicolumn{8}{|c|}{Subject 1 (more functional)} & \multicolumn{8}{|c|}{Subject 2 (less functional)}  \\ \hline
        ~ 
        & \begin{tabular}{@{}c@{}} Leg \\ func. \\ RMSE \end{tabular} 
        & \begin{tabular}{@{}c@{}} Rank \\ (left) \end{tabular}
        & \begin{tabular}{@{}c@{}} Leg \\ func. \\ CC \end{tabular}
        & \begin{tabular}{@{}c@{}} Rank \\ (left) \end{tabular}
        & \begin{tabular}{@{}c@{}} Leg \\ pare. \\ RMSE \end{tabular}
        & \begin{tabular}{@{}c@{}} Rank \\ (left) \end{tabular}
        & \begin{tabular}{@{}c@{}} Leg \\ pare. \\ CC \end{tabular}
        & \begin{tabular}{@{}c@{}} Rank \\ (left) \end{tabular}
        & \begin{tabular}{@{}c@{}} Leg \\ func. \\ RMSE \end{tabular}
        & \begin{tabular}{@{}c@{}} Rank \\ (left) \end{tabular}
        & \begin{tabular}{@{}c@{}} Leg \\ func. \\ CC \end{tabular}
        & \begin{tabular}{@{}c@{}} Rank \\ (left) \end{tabular}
        & \begin{tabular}{@{}c@{}} Leg \\ pare. \\ RMSE \end{tabular}
        & \begin{tabular}{@{}c@{}} Rank \\ (left) \end{tabular}
        & \begin{tabular}{@{}c@{}} Leg \\ pare. \\ CC \end{tabular}
        & \begin{tabular}{@{}c@{}} Rank \\ (left) \end{tabular}
        \\ \hline
        $\phi_1$ & 258 & 15 & 0.31 & 16 & 268 & 13 & 0.38 & 11 & 296 & 12 & 0.71 & 12 & 359 & 16 & 0.11 & 16 \\ \hline
        $\phi_2$ & 218 & 12 & 0.48 & 13 & 238 & 4 & 0.45 & 4 & 309 & 14 & 0.66 & 14 & 288 & 8 & 0.39 & 9 \\ \hline
        $\phi_3$ & 215 & 10 & 0.50 & 12 & 236 & 3 & 0.45 & 3 & 311 & 15 & 0.65 & 15 & 277 & 5 & 0.47 & 5 \\ \hline
        $\phi_4$ & 213 & 8 & 0.50 & 11 & 236 & 2 & 0.44 & 6 & 312 & 16 & 0.64 & 16 & 264 & 4 & 0.55 & 4 \\ \hline
        $\phi_5$ & 236 & 14 & 0.53 & 9 & 281 & 14 & 0.35 & 13 & 288 & 11 & 0.73 & 11 & 326 & 14 & 0.31 & 11 \\ \hline
        $\phi_6$ & 187 & 3 & 0.66 & 3 & 244 & 6 & 0.45 & 5 & 208 & 2 & 0.87 & 2 & 289 & 9 & 0.40 & 8 \\ \hline
        $\phi_7$ & 187 & 4 & 0.66 & 4 & 240 & 5 & 0.45 & 2 & 214 & 3 & 0.87 & 3 & 283 & 7 & 0.43 & 7 \\ \hline
        $\phi_8$ & 214 & 9 & 0.57 & 8 & 264 & 11 & 0.39 & 10 & 234 & 6 & 0.82 & 8 & 319 & 13 & 0.28 & 13 \\ \hline
        $\phi_9$ & 261 & 16 & 0.40 & 15 & 303 & 16 & 0.24 & 17 & 304 & 13 & 0.70 & 13 & 361 & 17 & 0.11 & 17 \\ \hline
        $\phi_{10}$ & 203 & 7 & 0.58 & 7 & 253 & 10 & 0.39 & 9 & 236 & 7 & 0.83 & 7 & 314 & 12 & 0.23 & 14 \\ \hline
        $\phi_{11}$ & 195 & 6 & 0.62 & 6 & 245 & 8 & 0.42 & 7 & 231 & 5 & 0.84 & 5 & 301 & 11 & 0.31 & 12 \\ \hline
        $\phi_{12}$ & 190 & 5 & 0.64 & 5 & 244 & 7 & 0.42 & 8 & 236 & 8 & 0.83 & 6 & 290 & 10 & 0.37 & 10 \\ \hline
        $\phi_{13}$ & 276 & 17 & 0.29 & 17 & 303 & 17 & 0.27 & 16 & 386 & 17 & 0.48 & 17 & 181 & 2 & 0.83 & 2 \\ \hline
        $\phi_{14}$ & 216 & 11 & 0.52 & 10 & 266 & 12 & 0.35 & 14 & 275 & 10 & 0.74 & 10 & 327 & 15 & 0.16 & 15 \\ \hline
        $\phi_{15}$ & 223 & 13 & 0.44 & 14 & 253 & 9 & 0.36 & 12 & 270 & 9 & 0.76 & 9 & 280 & 6 & 0.44 & 6 \\ \hline
        \rule{0pt}{15pt}
        $\Phi^{\rm S1, LEG1}_{\rm IOC}$ & 178 & 1 & 0.71 & 1 & ~ & ~ & ~ & ~ & 216 & 4 & 0.86 & 4 & ~ & ~ & ~ & ~ \\ \hline
        \rule{0pt}{15pt}
        $\Phi^{\rm S2, LEG1}_{\rm IOC}$ & 184 & 2 & 0.68 & 2 & ~ & ~ & ~ & ~ & 205 & 1 & 0.88 & 1 & ~ & ~ & ~ & ~ \\ \hline
        \rule{0pt}{15pt}
        $\Phi^{\rm S1, LEG2}_{\rm IOC}$ & ~ & ~ & ~ & ~ & 213 & 1 & 0.61 & 1 & ~ & ~ & ~ & ~ & 257 & 3 & 0.57 & 3 \\ \hline
        \rule{0pt}{15pt}
        $\Phi^{\rm S2, LEG2}_{\rm IOC}$ & ~ & ~ & ~ & ~ & 283 & 15 & 0.31 & 15 & ~ & ~ & ~ & ~ & 165 & 1 & 0.85 & 1 \\ \hline
    \end{tabular}
\end{sidewaystable}

\renewcommand\figurename{Supplementary Figure}
\renewcommand\thefigure{\Alph{figure}}
\setcounter{figure}{0}

\begin{sidewaysfigure}[!ht]
    \label{fig:force-trajectories}
    \caption{}
    \centering
    \includegraphics[width=\linewidth]{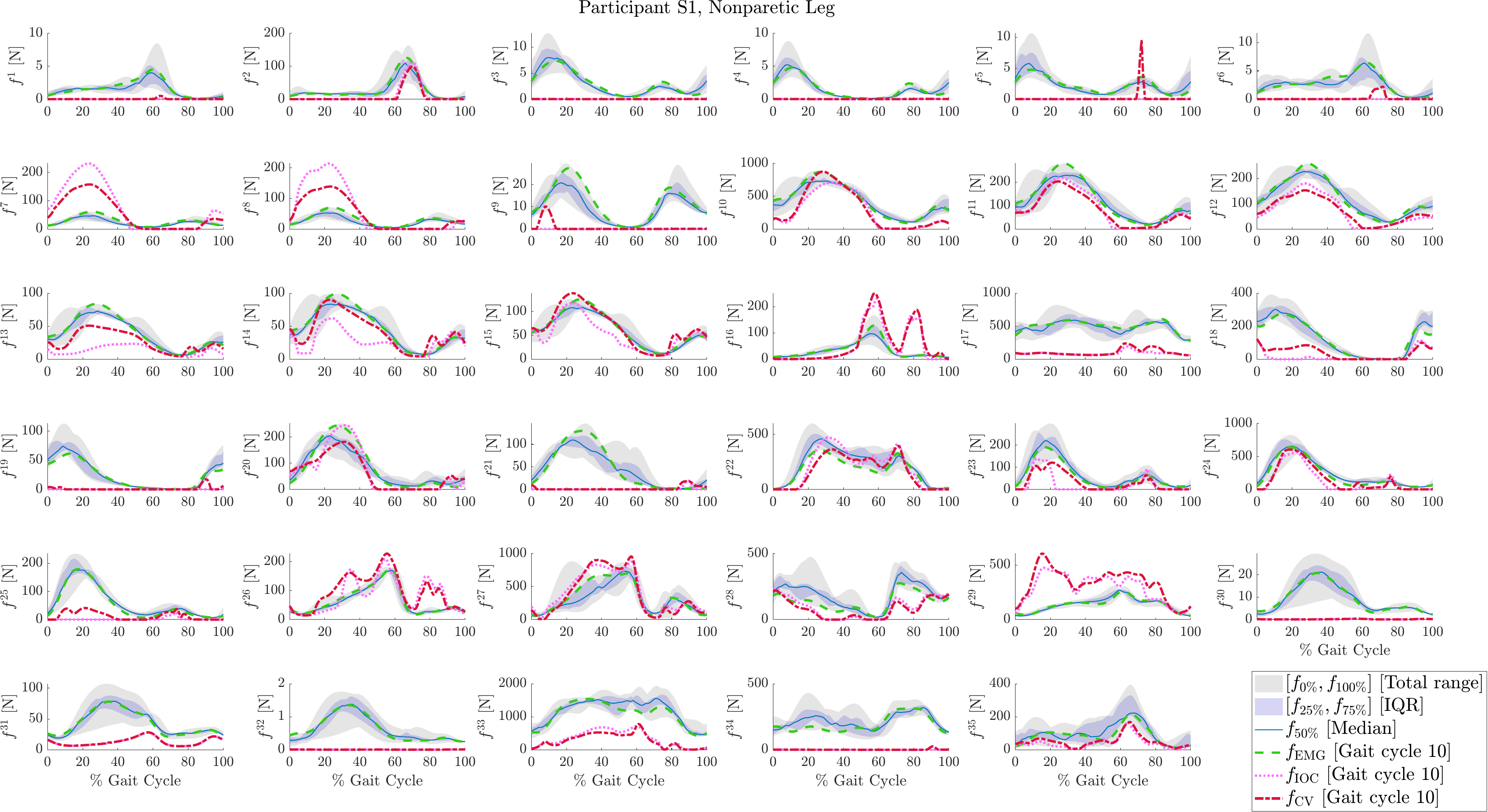}
\end{sidewaysfigure}

\clearpage
\section*{Acknowledgments}

\noindent 
F. Be\v{c}anovi\'{c} was financially supported by the Ministry of Science, Technological Development and Innovation of the Republic of Serbia under contract number: 451-03-136/2025-03/200103.
F. Be\v{c}anovi\'{c} and K. Jovanovi\'{c} were financially supported by the European Union’s Horizon Europe Programme for Research and Innovation through the MUSAE and CITADELS projects with grant agreement N\textsuperscript{o} 101070421 and N\textsuperscript{o} 101217281.
The research was partially conducted in the premises of the Palace of Science, Miodrag Kostić Endowment.

\bibliographystyle{elsarticle-harv} 
\bibliography{references}

\begin{thebibliography}{31}
\expandafter\ifx\csname natexlab\endcsname\relax\def\natexlab#1{#1}\fi
\providecommand{\url}[1]{\texttt{#1}}
\providecommand{\href}[2]{#2}
\providecommand{\path}[1]{#1}
\providecommand{\DOIprefix}{doi:}
\providecommand{\ArXivprefix}{arXiv:}
\providecommand{\URLprefix}{URL: }
\providecommand{\Pubmedprefix}{pmid:}
\providecommand{\doi}[1]{\href{http://dx.doi.org/#1}{\path{#1}}}
\providecommand{\Pubmed}[1]{\href{pmid:#1}{\path{#1}}}
\providecommand{\bibinfo}[2]{#2}
\ifx\xfnm\relax \def\xfnm[#1]{\unskip,\space#1}\fi
\bibitem[{Ackermann and Van~den Bogert(2010)}]{ackermann2010optimality}
\bibinfo{author}{Ackermann, M.}, \bibinfo{author}{Van~den Bogert, A.J.},
  \bibinfo{year}{2010}.
\newblock \bibinfo{title}{Optimality principles for model-based prediction of
  human gait}.
\newblock \bibinfo{journal}{Journal of biomechanics} \bibinfo{volume}{43},
  \bibinfo{pages}{1055--1060}.
\bibitem[{Andersson et~al.(2018)Andersson, Gillis, Horn, Rawlings and
  Diehl}]{Andersson2018}
\bibinfo{author}{Andersson, J.A.E.}, \bibinfo{author}{Gillis, J.},
  \bibinfo{author}{Horn, G.}, \bibinfo{author}{Rawlings, J.B.},
  \bibinfo{author}{Diehl, M.}, \bibinfo{year}{2018}.
\newblock \bibinfo{title}{{CasADi} -- {A} software framework for nonlinear
  optimization and optimal control}.
\newblock \bibinfo{journal}{Mathematical Programming Computation} .
\bibitem[{Barati et~al.(2023)Barati, Nazari, Kardan and
  Akbarzadeh}]{barati2023predictive}
\bibinfo{author}{Barati, H.}, \bibinfo{author}{Nazari, K.},
  \bibinfo{author}{Kardan, I.}, \bibinfo{author}{Akbarzadeh, A.},
  \bibinfo{year}{2023}.
\newblock \bibinfo{title}{Predictive simulation of hemiparetic gait based on
  neural impairments}, in: \bibinfo{booktitle}{2023 11th RSI International
  Conference on Robotics and Mechatronics (ICRoM)},
  \bibinfo{organization}{IEEE}. pp. \bibinfo{pages}{371--376}.
\bibitem[{Bersani et~al.(2023)Bersani, Davico and
  Viceconti}]{bersani2023modeling}
\bibinfo{author}{Bersani, A.}, \bibinfo{author}{Davico, G.},
  \bibinfo{author}{Viceconti, M.}, \bibinfo{year}{2023}.
\newblock \bibinfo{title}{Modeling human suboptimal control: A review}.
\newblock \bibinfo{journal}{Journal of Applied Biomechanics}
  \bibinfo{volume}{39}, \bibinfo{pages}{294--303}.
\bibitem[{Bečanović et~al.(2023)Bečanović, Bonnet, Dumas, Jovanović and
  Mohammed}]{bevcanovic2023force}
\bibinfo{author}{Bečanović, F.}, \bibinfo{author}{Bonnet, V.},
  \bibinfo{author}{Dumas, R.}, \bibinfo{author}{Jovanović, K.},
  \bibinfo{author}{Mohammed, S.}, \bibinfo{year}{2023}.
\newblock \bibinfo{title}{Force sharing problem during gait using inverse
  optimal control}.
\newblock \bibinfo{journal}{IEEE Robotics and Automation Letters}
  \bibinfo{volume}{8}, \bibinfo{pages}{872--879}.
\newblock \DOIprefix\doi{10.1109/LRA.2022.3217398}.
\bibitem[{Boyd and Vandenberghe(2004)}]{boyd2004convex}
\bibinfo{author}{Boyd, S.P.}, \bibinfo{author}{Vandenberghe, L.},
  \bibinfo{year}{2004}.
\newblock \bibinfo{title}{Convex optimization}.
\newblock \bibinfo{publisher}{Cambridge university press}.
\bibitem[{Chan et~al.(2025)Chan, Mahmood and Zhu}]{chan2025inverse}
\bibinfo{author}{Chan, T.C.}, \bibinfo{author}{Mahmood, R.},
  \bibinfo{author}{Zhu, I.Y.}, \bibinfo{year}{2025}.
\newblock \bibinfo{title}{Inverse optimization: Theory and applications}.
\newblock \bibinfo{journal}{Operations Research} \bibinfo{volume}{73},
  \bibinfo{pages}{1046--1074}.
\bibitem[{De~Groote et~al.(2016)De~Groote, Kinney, Rao and
  Fregly}]{de2016evaluation}
\bibinfo{author}{De~Groote, F.}, \bibinfo{author}{Kinney, A.L.},
  \bibinfo{author}{Rao, A.V.}, \bibinfo{author}{Fregly, B.J.},
  \bibinfo{year}{2016}.
\newblock \bibinfo{title}{Evaluation of direct collocation optimal control
  problem formulations for solving the muscle redundancy problem}.
\newblock \bibinfo{journal}{Annals of biomedical engineering}
  \bibinfo{volume}{44}, \bibinfo{pages}{2922--2936}.
\bibitem[{Denayer et~al.(2025)Denayer, Alfio, D{\'\i}az, Sartori, De~Groote,
  De~Pauw and Verstraten}]{denayer2025prisma}
\bibinfo{author}{Denayer, M.}, \bibinfo{author}{Alfio, E.},
  \bibinfo{author}{D{\'\i}az, M.A.}, \bibinfo{author}{Sartori, M.},
  \bibinfo{author}{De~Groote, F.}, \bibinfo{author}{De~Pauw, K.},
  \bibinfo{author}{Verstraten, T.}, \bibinfo{year}{2025}.
\newblock \bibinfo{title}{A prisma systematic review through time on predictive
  musculoskeletal simulations}.
\newblock \bibinfo{journal}{Journal of NeuroEngineering and Rehabilitation}
  \bibinfo{volume}{22}, \bibinfo{pages}{1--24}.
\bibitem[{Erdemir et~al.(2007)Erdemir, McLean, Herzog and van~den
  Bogert}]{erdemir2007model}
\bibinfo{author}{Erdemir, A.}, \bibinfo{author}{McLean, S.},
  \bibinfo{author}{Herzog, W.}, \bibinfo{author}{van~den Bogert, A.J.},
  \bibinfo{year}{2007}.
\newblock \bibinfo{title}{Model-based estimation of muscle forces exerted
  during movements}.
\newblock \bibinfo{journal}{Clinical biomechanics} \bibinfo{volume}{22},
  \bibinfo{pages}{131--154}.
\bibitem[{Falisse et~al.(2020)Falisse, Pitto, Kainz, Hoang, Wesseling,
  Van~Rossom, Papageorgiou, Bar-On, Hallemans, Desloovere
  et~al.}]{falisse2020physics}
\bibinfo{author}{Falisse, A.}, \bibinfo{author}{Pitto, L.},
  \bibinfo{author}{Kainz, H.}, \bibinfo{author}{Hoang, H.},
  \bibinfo{author}{Wesseling, M.}, \bibinfo{author}{Van~Rossom, S.},
  \bibinfo{author}{Papageorgiou, E.}, \bibinfo{author}{Bar-On, L.},
  \bibinfo{author}{Hallemans, A.}, \bibinfo{author}{Desloovere, K.}, et~al.,
  \bibinfo{year}{2020}.
\newblock \bibinfo{title}{Physics-based simulations to predict the differential
  effects of motor control and musculoskeletal deficits on gait dysfunction in
  cerebral palsy: a retrospective case study}.
\newblock \bibinfo{journal}{Frontiers in human neuroscience}
  \bibinfo{volume}{14}, \bibinfo{pages}{40}.
\bibitem[{Febrer-Nafr{\'{i}}a et~al.(2023)Febrer-Nafr{\'{i}}a, Nasr, Ezati,
  Brown, Font-Llagunes and McPhee}]{febrer2023predictive}
\bibinfo{author}{Febrer-Nafr{\'{i}}a, M.}, \bibinfo{author}{Nasr, A.},
  \bibinfo{author}{Ezati, M.}, \bibinfo{author}{Brown, P.},
  \bibinfo{author}{Font-Llagunes, J.M.}, \bibinfo{author}{McPhee, J.},
  \bibinfo{year}{2023}.
\newblock \bibinfo{title}{Predictive multibody dynamic simulation of human
  neuromusculoskeletal systems: a review}.
\newblock \bibinfo{journal}{Multibody System Dynamics} \bibinfo{volume}{58},
  \bibinfo{pages}{299--339}.
\bibitem[{Hammond et~al.(2025)Hammond, Williams, Vega, Ao, Li, Salati, Pariser,
  Shourijeh, Habib, Patten et~al.}]{hammond2025neuromusculoskeletal}
\bibinfo{author}{Hammond, C.V.}, \bibinfo{author}{Williams, S.T.},
  \bibinfo{author}{Vega, M.M.}, \bibinfo{author}{Ao, D.}, \bibinfo{author}{Li,
  G.}, \bibinfo{author}{Salati, R.M.}, \bibinfo{author}{Pariser, K.M.},
  \bibinfo{author}{Shourijeh, M.S.}, \bibinfo{author}{Habib, A.W.},
  \bibinfo{author}{Patten, C.}, et~al., \bibinfo{year}{2025}.
\newblock \bibinfo{title}{The neuromusculoskeletal modeling pipeline:
  Matlab-based model personalization and treatment optimization functionality
  for opensim}.
\newblock \bibinfo{journal}{Journal of NeuroEngineering and Rehabilitation}
  \bibinfo{volume}{22}, \bibinfo{pages}{112}.
\bibitem[{Hu et~al.(2013)Hu, Lu and Chen}]{hu2013influence}
\bibinfo{author}{Hu, C.C.}, \bibinfo{author}{Lu, T.W.}, \bibinfo{author}{Chen,
  S.C.}, \bibinfo{year}{2013}.
\newblock \bibinfo{title}{Influence of model complexity and problem formulation
  on the forces in the knee calculated using optimization methods}.
\newblock \bibinfo{journal}{Biomedical engineering online}
  \bibinfo{volume}{12}, \bibinfo{pages}{1--16}.
\bibitem[{Jansen et~al.(2014)Jansen, De~Groote, Aerts, De~Schutter, Duysens and
  Jonkers}]{jansen2014altering}
\bibinfo{author}{Jansen, K.}, \bibinfo{author}{De~Groote, F.},
  \bibinfo{author}{Aerts, W.}, \bibinfo{author}{De~Schutter, J.},
  \bibinfo{author}{Duysens, J.}, \bibinfo{author}{Jonkers, I.},
  \bibinfo{year}{2014}.
\newblock \bibinfo{title}{Altering length and velocity feedback during a
  neuro-musculoskeletal simulation of normal gait contributes to hemiparetic
  gait characteristics}.
\newblock \bibinfo{journal}{Journal of neuroengineering and rehabilitation}
  \bibinfo{volume}{11}, \bibinfo{pages}{1--15}.
\bibitem[{Johnson et~al.(2022)Johnson, Bianco and Finley}]{johnson2022patterns}
\bibinfo{author}{Johnson, R.T.}, \bibinfo{author}{Bianco, N.A.},
  \bibinfo{author}{Finley, J.M.}, \bibinfo{year}{2022}.
\newblock \bibinfo{title}{Patterns of asymmetry and energy cost generated from
  predictive simulations of hemiparetic gait}.
\newblock \bibinfo{journal}{PLoS computational biology} \bibinfo{volume}{18},
  \bibinfo{pages}{e1010466}.
\bibitem[{Li et~al.(2021)Li, Shourijeh, Ao, Patten and Fregly}]{li2021well}
\bibinfo{author}{Li, G.}, \bibinfo{author}{Shourijeh, M.S.},
  \bibinfo{author}{Ao, D.}, \bibinfo{author}{Patten, C.},
  \bibinfo{author}{Fregly, B.J.}, \bibinfo{year}{2021}.
\newblock \bibinfo{title}{How well do commonly used co-contraction indices
  approximate lower limb joint stiffness trends during gait for individuals
  post-stroke?}
\newblock \bibinfo{journal}{Frontiers in Bioengineering and Biotechnology} ,
  \bibinfo{pages}{1503}\bibinfo{note}{Data available at:
  \hyperlink{https://simtk.org/projects/ccivsjointstiff}{https://simtk.org/projects/ccivsjointstiff}}.
\bibitem[{Meyer et~al.(2017)Meyer, Patten and Fregly}]{meyer2017lower}
\bibinfo{author}{Meyer, A.J.}, \bibinfo{author}{Patten, C.},
  \bibinfo{author}{Fregly, B.J.}, \bibinfo{year}{2017}.
\newblock \bibinfo{title}{Lower extremity emg-driven modeling of walking with
  automated adjustment of musculoskeletal geometry}.
\newblock \bibinfo{journal}{PloS one} \bibinfo{volume}{12},
  \bibinfo{pages}{e0179698}.
\bibitem[{Modenese et~al.(2011)Modenese, Phillips and Bull}]{modenese2011open}
\bibinfo{author}{Modenese, L.}, \bibinfo{author}{Phillips, A.},
  \bibinfo{author}{Bull, A.}, \bibinfo{year}{2011}.
\newblock \bibinfo{title}{An open source lower limb model: Hip joint
  validation}.
\newblock \bibinfo{journal}{Journal of biomechanics} \bibinfo{volume}{44},
  \bibinfo{pages}{2185--2193}.
\bibitem[{Moissenet et~al.(2014)Moissenet, Cheze and Dumas}]{moissenet20143d}
\bibinfo{author}{Moissenet, F.}, \bibinfo{author}{Cheze, L.},
  \bibinfo{author}{Dumas, R.}, \bibinfo{year}{2014}.
\newblock \bibinfo{title}{A 3d lower limb musculoskeletal model for
  simultaneous estimation of musculo-tendon, joint contact, ligament and bone
  forces during gait}.
\newblock \bibinfo{journal}{Journal of biomechanics} \bibinfo{volume}{47},
  \bibinfo{pages}{50--58}.
\bibitem[{Moissenet et~al.(2013)Moissenet, Pradon, Lampire, Dumas and
  Cheze}]{moissenet2013new}
\bibinfo{author}{Moissenet, F.}, \bibinfo{author}{Pradon, D.},
  \bibinfo{author}{Lampire, N.}, \bibinfo{author}{Dumas, R.},
  \bibinfo{author}{Cheze, L.}, \bibinfo{year}{2013}.
\newblock \bibinfo{title}{A new optimization criterion introducing the muscle
  stretch velocity in the muscular redundancy problem: A first step into the
  modeling of spastic muscle}.
\newblock \bibinfo{journal}{Modeling, Simulation and Optimization of Bipedal
  Walking} , \bibinfo{pages}{155--164}.
\bibitem[{Nguyen et~al.(2019)Nguyen, Johnson, Sup and
  Umberger}]{nguyen2019bilevel}
\bibinfo{author}{Nguyen, V.Q.}, \bibinfo{author}{Johnson, R.T.},
  \bibinfo{author}{Sup, F.C.}, \bibinfo{author}{Umberger, B.R.},
  \bibinfo{year}{2019}.
\newblock \bibinfo{title}{Bilevel optimization for cost function determination
  in dynamic simulation of human gait}.
\newblock \bibinfo{journal}{IEEE Transactions on Neural Systems and
  Rehabilitation Engineering} \bibinfo{volume}{27},
  \bibinfo{pages}{1426--1435}.
\bibitem[{Praagman et~al.(2006)Praagman, Chadwick, Van Der~Helm and
  Veeger}]{praagman2006relationship}
\bibinfo{author}{Praagman, M.}, \bibinfo{author}{Chadwick, E.},
  \bibinfo{author}{Van Der~Helm, F.}, \bibinfo{author}{Veeger, H.},
  \bibinfo{year}{2006}.
\newblock \bibinfo{title}{The relationship between two different mechanical
  cost functions and muscle oxygen consumption}.
\newblock \bibinfo{journal}{Journal of biomechanics} \bibinfo{volume}{39},
  \bibinfo{pages}{758--765}.
\bibitem[{Prilutsky and Zatsiorsky(2002)}]{prilutsky2002optimization}
\bibinfo{author}{Prilutsky, B.I.}, \bibinfo{author}{Zatsiorsky, V.M.},
  \bibinfo{year}{2002}.
\newblock \bibinfo{title}{Optimization-based models of muscle coordination}.
\newblock \bibinfo{journal}{Exercise and sport sciences reviews}
  \bibinfo{volume}{30}, \bibinfo{pages}{32--38}.
\bibitem[{Rasmussen et~al.(2001)Rasmussen, Damsgaard and
  Voigt}]{rasmussen2001muscle}
\bibinfo{author}{Rasmussen, J.}, \bibinfo{author}{Damsgaard, M.},
  \bibinfo{author}{Voigt, M.}, \bibinfo{year}{2001}.
\newblock \bibinfo{title}{Muscle recruitment by the min/max criterion—a
  comparative numerical study}.
\newblock \bibinfo{journal}{Journal of biomechanics} \bibinfo{volume}{34},
  \bibinfo{pages}{409--415}.
\bibitem[{Serrancol{\'\i} et~al.(2014)Serrancol{\'\i}, Font-Llagunes and
  Barjau}]{serrancoli2014weighted}
\bibinfo{author}{Serrancol{\'\i}, G.}, \bibinfo{author}{Font-Llagunes, J.M.},
  \bibinfo{author}{Barjau, A.}, \bibinfo{year}{2014}.
\newblock \bibinfo{title}{A weighted cost function to deal with the muscle
  force sharing problem in injured subjects: A single case study}.
\newblock \bibinfo{journal}{Proceedings of the Institution of Mechanical
  Engineers, Part K: Journal of Multi-body Dynamics} \bibinfo{volume}{228},
  \bibinfo{pages}{241--251}.
\bibitem[{Tomasi and Artoni(2023)}]{tomasi2023identification}
\bibinfo{author}{Tomasi, M.}, \bibinfo{author}{Artoni, A.},
  \bibinfo{year}{2023}.
\newblock \bibinfo{title}{Identification of motor control objectives in human
  locomotion via multi-objective inverse optimal control}.
\newblock \bibinfo{journal}{Journal of Computational and Nonlinear Dynamics}
  \bibinfo{volume}{18}, \bibinfo{pages}{051004}.
\bibitem[{Tsirakos et~al.(1997)Tsirakos, Baltzopoulos and
  Bartlett}]{tsirakos1997inverse}
\bibinfo{author}{Tsirakos, D.}, \bibinfo{author}{Baltzopoulos, V.},
  \bibinfo{author}{Bartlett, R.}, \bibinfo{year}{1997}.
\newblock \bibinfo{title}{Inverse optimization: functional and physiological
  considerations related to the force-sharing problem}.
\newblock \bibinfo{journal}{Critical Reviews™ in Biomedical Engineering}
  \bibinfo{volume}{25}.
\bibitem[{Veerkamp et~al.(2023)Veerkamp, Carty, Waterval, Geijtenbeek, Buizer,
  Lloyd, Harlaar and van~der Krogt}]{veerkamp2023predicting}
\bibinfo{author}{Veerkamp, K.}, \bibinfo{author}{Carty, C.P.},
  \bibinfo{author}{Waterval, N.F.}, \bibinfo{author}{Geijtenbeek, T.},
  \bibinfo{author}{Buizer, A.I.}, \bibinfo{author}{Lloyd, D.G.},
  \bibinfo{author}{Harlaar, J.}, \bibinfo{author}{van~der Krogt, M.M.},
  \bibinfo{year}{2023}.
\newblock \bibinfo{title}{Predicting gait patterns of children with spasticity
  by simulating hyperreflexia}.
\newblock \bibinfo{journal}{Journal of applied biomechanics}
  \bibinfo{volume}{1}, \bibinfo{pages}{1--13}.
\bibitem[{Veerkamp et~al.(2019)Veerkamp, Schallig, Harlaar, Pizzolato, Carty,
  Lloyd and van~der Krogt}]{veerkamp2019effects}
\bibinfo{author}{Veerkamp, K.}, \bibinfo{author}{Schallig, W.},
  \bibinfo{author}{Harlaar, J.}, \bibinfo{author}{Pizzolato, C.},
  \bibinfo{author}{Carty, C.P.}, \bibinfo{author}{Lloyd, D.G.},
  \bibinfo{author}{van~der Krogt, M.M.}, \bibinfo{year}{2019}.
\newblock \bibinfo{title}{The effects of electromyography-assisted modelling in
  estimating musculotendon forces during gait in children with cerebral palsy}.
\newblock \bibinfo{journal}{Journal of biomechanics} \bibinfo{volume}{92},
  \bibinfo{pages}{45--53}.
\bibitem[{Zargham et~al.(2019)Zargham, Afschrift, De~Schutter, Jonkers and
  De~Groote}]{zargham2019inverse}
\bibinfo{author}{Zargham, A.}, \bibinfo{author}{Afschrift, M.},
  \bibinfo{author}{De~Schutter, J.}, \bibinfo{author}{Jonkers, I.},
  \bibinfo{author}{De~Groote, F.}, \bibinfo{year}{2019}.
\newblock \bibinfo{title}{Inverse dynamic estimates of muscle recruitment and
  joint contact forces are more realistic when minimizing muscle activity
  rather than metabolic energy or contact forces}.
\newblock \bibinfo{journal}{Gait \& Posture} \bibinfo{volume}{74},
  \bibinfo{pages}{223--230}.

\end{thebibliography}
\end{document}